\documentclass[11pt,prd,reprint,notitlepage,nofootinbib]{revtex4-1}

\usepackage{amsmath}
\usepackage{amsfonts}
\usepackage{graphicx}
\usepackage{aas_macros}
\usepackage{color}
\usepackage{listings}

\newcommand{\mv}[1]{}

\begin{document}

\title{Taming outliers in pulsar-timing datasets with hierarchical likelihoods and Hamiltonian sampling}
\author{Michele Vallisneri and Rutger van Haasteren}
\address{Jet Propulsion Laboratory, California Institute of Technology, Pasadena, California 91109}

\begin{abstract}
Pulsar-timing datasets have been analyzed with great success using probabilistic treatments based on Gaussian distributions, with applications ranging from studies of neutron-star structure to tests of general relativity and searches for nanosecond gravitational waves. 
As for other applications of Gaussian distributions, \emph{outliers} in timing measurements pose a significant challenge to statistical inference, since they can bias the estimation of timing and noise parameters, and affect reported parameter uncertainties.
We describe and demonstrate a practical end-to-end approach to perform Bayesian inference of timing and noise parameters \emph{robustly} in the presence of outliers, and to identify these probabilistically. 
The method is fully consistent (i.e., outlier-ness probabilities vary in tune with the posterior distributions of the timing and noise parameters), and it relies on the efficient sampling of the hierarchical form of the pulsar-timing likelihood. Such sampling has recently become possible with a ``no-U-turn'' Hamiltonian sampler coupled to a highly customized reparametrization of the likelihood;
this code is described elsewhere, but it is already available online.
We recommend our method as a standard step in the preparation of pulsar-timing-array datasets: even if statistical inference is not affected, follow-up studies of outlier candidates can reveal unseen problems in radio observations and timing measurements; furthermore, confidence in the results of gravitational-wave searches will only benefit from stringent statistical evidence that datasets are clean and outlier-free.
\end{abstract}

\maketitle

\paragraph*{Pulsar timing.} 
The scientific value of pulsar timing \cite{2012hpa..book.....L} lies in the possibility of performing very accurate fits of very detailed physical models, allowing remarkable applications and discoveries, such as characterizing the structure and physics of pulsars \cite{2007PhR...442..109L}\mv{anything better?}, testing general relativity \cite{2003LRR.....6....5S}, identifying extrasolar planets \cite{1992Natur.355..145W}, mapping free-electron density across the Galaxy \cite{2002astro.ph..7156C,2003astro.ph..1598C}, and (of course) searching for gravitational waves. See \cite{2015RPPh...78l4901L,2015arXiv151107869B} for recent reviews.
Here we use ``fit'' as a loose term for ``statistical inference'' (whether of the Bayesian or classical variety), whereby a probabilistic model of the noise is used with the data to derive estimates for the timing parameters of the pulsar. The noise model may incorporate components due to the timing measurements, to intrinsic irregularities in the periodic emission of pulses, and to delays induced through propagation in the interstellar medium. The timing parameters comprise a basic rotation model, astrometric parameters, and orbital elements for binary pulsars \cite{2013CQGra..30v4001L,2006MNRAS.372.1549E}\mv{review these refs}.
We may even include deterministically modeled or noise-like gravitational waves, and endeavor to establish their presence or to limit their amplitude (most recently: \cite{2015MNRAS.453.2576L,2015Sci...349.1522S,2016ApJ...821...13A}), alongside our estimation of timing parameters.
 
Mathematically, probabilistic models of noise are almost always based on Gaussian distributions (however, see \cite{2014MNRAS.444.3863L}). The ``radiometer'' errors incurred in measuring individual pulse times-of-arrival (TOAs) are taken as independent normal variables, each with a different variance, a function of the signal-to-noise ratio (SNR) of each observation. Even time-correlated pulsar-spin noise can be described as a Gaussian process (see \cite{2014PhRvD..90j4012V} for a recent review). This leads to a likelihood -- the probability of obtaining the observed data as a function of the timing-model and noise parameters -- in the form of a joint Gaussian distribution. For such a likelihood, the inference problem can be solved analytically if the noise parameters are fixed and the effect of the timing parameters is linearized, or at least the problem can be attacked numerically with surprising efficiency \cite{2014PhRvD..90j4012V}.

\paragraph*{The problem with outliers.}
Unfortunately, Gaussian likelihoods are very vulnerable to the presence of \emph{outliers} in the dataset. These are datapoints that have a physical origin other than the process reflected by our deterministic/probabilistic model of the data. For instance, outliers may arise in low-SNR timing observations from strong thermal-noise spikes being mistaken for actual radio pulses; in this case, not only are the outliers spread much more broadly
than ``good'' measurements, but they do not even center around the true TOAs. More generally, the statistical distribution of outliers can be very different (biased, much broader, and non-Gaussian) than represented in our formulas, affecting the accuracy of statistical inference to a degree that depends on the number and severity of the outliers. Using the jargon of least-squares fitting (appropriate for independently distributed errors), we may say that outliers are displaced by several sigmas from the best fit that could be derived if the outliers were not in the data; thus, the outliers disproportionately affect the chi square, a quadratic function of the ``sigma values,'' and may end up dominating a fit that includes them.
\begin{figure*}
\includegraphics[width=\textwidth]{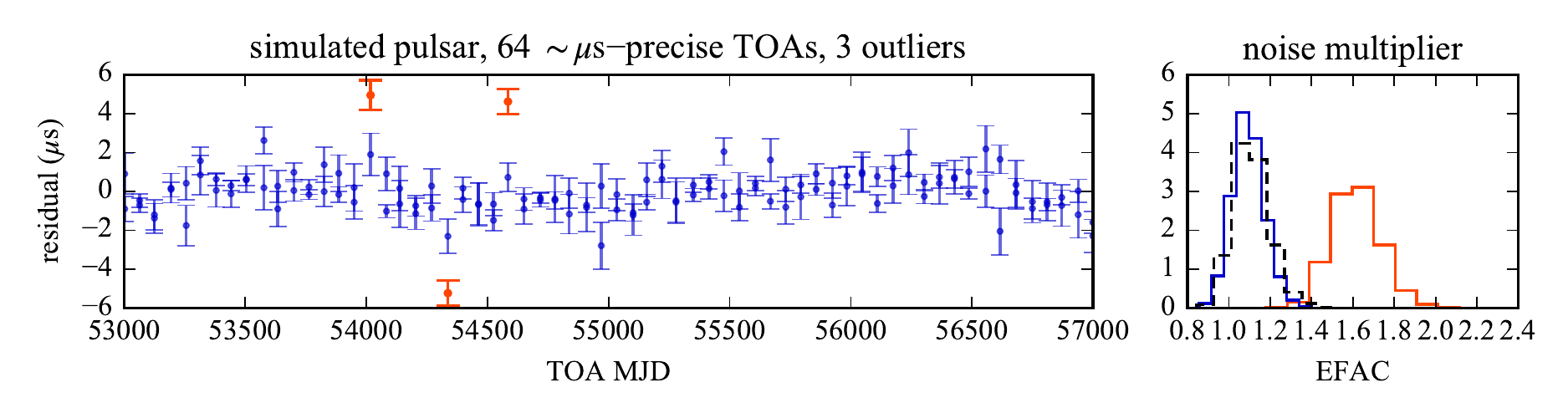}
\vspace{-24pt}
\caption{Left: simulated pulsar-timing dataset with the addition of three strong outliers (thick red markers). We show residuals computed against the best-fit timing-model and noise parameters. Right: posterior probability distribution for the EFAC noise multiplier, as computed in the presence of outliers (red histogram, displaced to EFAC $\simeq 1.6$), after excluding them outright (blue histogram), and with the outlier-robust analysis described later in this paper (dashed black histogram). The analysis identifies all three outliers correctly, with $P_{i,\mathrm{out}} \simeq 1$. The dataset was generated using the \texttt{libstempo} Python interface to \texttt{tempo2} (\texttt{github.com/vallis/libstempo}) and the \texttt{libstempo/toasim} module, on the basis of the PSR J0030+0451 timing parameters (with simplifications).\label{fig:simulated}}
\end{figure*}

To make this discussion more concrete for the case of pulsar timing, in Fig.\ \ref{fig:simulated} we show a simulated dataset of 64 TOAs (based on PSR J0030+0451, with significant simplifications in the timing model). We have introduced three large outliers, identified by the thick red dots and errorbars in the left panel of the figure.
The outliers bias the estimation of the TOA measurement noise: as shown by the red profile in the right panel, the Bayesian posterior probability for the ``EFAC'' parameter (which multiplies individual measurement errors in the dataset) is displaced to 1.6$\times$ the correct value of 1, to account for the additional outlier-induced variance. By contrast, the blue profile in the right panel shows the estimate of noise that would be obtained if the outliers were not present.
In this fit, the effect of the outliers on the timing-model parameters is only to increase their uncertainty (since the fit prefers more measurement noise) rather than to bias their estimates---which can nevertheless happen, depending on the configuration of the outliers.

\paragraph*{Outlier mitigation.}
What to do? For data contamination as blatant as in Fig.\ \ref{fig:simulated}, we may just identify the outliers visually and exclude them, or at least inspect the original TOA measurements and look for anomalies. However, such a manual solution is incompatible with reproducibility and unbiasedness, and it is also impractical for large amounts of data. A variety of more objective outlier-mitigation algorithms have been proposed in the statistical literature \cite{1987rrod.book.....L,barnettlewis1994,hawkins2013}.
Perhaps the simplest approach, known as \emph{sigma clipping} (a variant of iterative deletion \cite{1987rrod.book.....L}), can be formulated as follows in the context of linear least-squares estimation. Let our problem be described by 
\begin{equation}
\label{eq:leastsquares}
y_i = \sum_{\mu=1}^P M_{i\mu} \eta_\mu + \epsilon_i, \quad
\text{for $i = 1,\ldots,N$};
\end{equation}
here the $y_i$ are the $N$ measurements; the $\eta_\mu$ are the $P$ parameters that we wish to estimate; $M_{i \mu}$ is the \emph{design matrix} (whose columns may encode, e.g., a constant, a linear trend, a quadratic); and the $\epsilon_i$ are (unknown) measurement errors taken to be independently distributed as Gaussians, $\epsilon_i \sim \mathcal{N}(0,\sigma^2_i)$---except that some are instead outliers drawn from a different, much broader distribution.

In sigma clipping, we first fit the model using all the data, resulting in the parameter estimates $\eta^{(0)}_\mu$; we compute the post-fit residuals $r^{(0)}_i = y_i - \sum_\mu M_{i\mu} \eta^{(0)}_\mu$; we identify the datapoints for which $r^{(0)}_i / \sigma_i$ is greater than a set threshold,
large enough that such an error would be very unlikely to appear in the data; we deem the worst offending point an outlier, and discard it from the dataset; we fit the model again, resulting in the updated (and hopefully less biased) parameter estimates $\eta^{(1)}_\mu$ and residuals $r^{(1)}_i$; and we continue iteratively until no residuals are found above the sigma threshold. At every step, we remove the datapoint that contributes the most to the fit's $\chi^2$, defined as $\sum_i r^2_i/\sigma^2_i$.

Sigma clipping is straightforward and makes intuitive sense, but it does not generalize well to the pulsar-timing case, for two reasons: first, because the noise parameters enter the computation of the likelihood, there is no unique set of residuals that may be used to define outlier-ness; second, in the presence of red timing noise or dispersion-measure fluctuations, the stochastic components of the TOAs become correlated and the likelihood has the form $\exp\{-\mathbf{r}^T \mathsf{C}^{-1}\mathbf{r}/2\}$, with $\mathsf{C}$ a dense matrix, so the contribution of each datapoint to $\chi^2$ cannot be isolated.

\paragraph*{A Bayesian mixture model of outliers.}
A statistically more principled procedure (advocated by Hogg, Bovy, and Lang \cite{2010arXiv1008.4686H}, and discussed more formally in \cite{1997upa..conf...49P,jaynes2003}) follows from recognizing that least-squares estimation is equivalent to maximizing the likelihood
\begin{widetext}
%
\begin{equation}
p(\mathbf{y}|\boldsymbol{\eta}) = 
\prod_i p(y_i|\boldsymbol{\eta}) = 
\prod_i \left[ \exp\Bigl\{-(y_i - \sum_\mu M_{i\mu} \eta_\mu)^2/(2 \sigma_i^2)\Bigr\} / \sqrt{2 \pi \sigma_i^2} \right]
\end{equation}
%
(which is indeed proportional to $\mathrm{e}^{-\chi^2/2}$), and from replacing the likelihood of each individual measurement with an expression that allows for the possibility that the measurement is an outlier: 
%
\begin{equation}
\label{eq:prime}
p'(y_j|b_i,\boldsymbol{\eta};\sigma_\mathrm{out}) 
= \left\{ \begin{array}{l@{\quad}l}
\mathrm{e}^{-(y_i - \sum_\mu M_{i\mu} \eta_\mu)^2/(2 \sigma_i^2)} / \sqrt{2 \pi \sigma_i^2} & \text{for $b_i = 0$,} \\[6pt]
\mathrm{e}^{-y_i^2/(2 \sigma_\mathrm{out}^2)} / \sqrt{2 \pi \sigma_\mathrm{out}^2} & \text{for $b_i = 1$;}
\end{array}\right.
\end{equation}
\end{widetext}
here the $b_i$ are binary labels that identify each $y_i$ as either a regular datapoint or an outlier, and $\sigma_\mathrm{out}$ (with $\sigma_\mathrm{out} \gg$ every $\sigma_i$) represents the typical range of outlier fluctuations. (Note that we are slightly modifying Hogg \emph{et al.}'s treatment by modeling outliers that are not just much noisier measurements, but measurements of \emph{noise alone}.) This likelihood can be maximized \emph{as is} to obtain the most likely model parameters $\eta_\mu$ and outlier classifications $b_i$.
In a Bayesian-inference context, if we provide a prior for the $b_i$,
\begin{equation}
P(b_i = 1) = P_\mathrm{out}, \quad P(b_i = 0) = 1 - P_\mathrm{out},
\end{equation}
we may also \emph{marginalize} Eq.\ \eqref{eq:pprime} with respect to the $b_i$, yielding a remarkably simple \emph{mixture} expression:
\begin{multline}
\label{eq:pprime}
p''(y_j|\boldsymbol{\eta};\sigma_\mathrm{out},P_\mathrm{out}) =
(1 - P_\mathrm{out}) \times p'(y_i|b_i=0,\boldsymbol{\eta}) \, + \\
P_\mathrm{out} \times p'(y_i|b_i=1,\boldsymbol{\eta};\sigma_\mathrm{out}),
\end{multline}
where $p'(y_i|b_i=0,\boldsymbol{\eta})$ and $p'(y_i|b_i=1,\boldsymbol{\eta};\sigma_\mathrm{out})$ are given by the two rows of Eq.\ \eqref{eq:prime}.
As it is manifest in Eq.\ \eqref{eq:pprime}, we are not completely excluding points that are exceedingly unlikely (as in sigma clipping), but instead we allow every point to behave as a regular measurement or an outlier, according to $P_\mathrm{out}$ and to the relative weight of $p'(y_i|b_i=0,\boldsymbol{\eta})$ and $p'(y_i|b_i=1,\boldsymbol{\eta};\sigma_\mathrm{out})$.

We can now perform statistical inference using the full-dataset likelihood $p''(\mathbf{y}|\boldsymbol{\eta};\sigma_\mathrm{out},P_\mathrm{out}) = \prod_i p''(y_i|\boldsymbol{\eta};\sigma_\mathrm{out},P_\mathrm{out})$, gaining robustness against outliers at the cost of adding the parameters $\sigma_\mathrm{out}$ and $P_\mathrm{out}$. (In fact, these are \emph{hyperparameters}, since they determine the form of the likelihood for the regular parameters $\eta_\mu$.) In Bayesian inference, we can hold $\sigma_\mathrm{out}$ and $P_\mathrm{out}$ fixed to reasonable values; or, more naturally, we can assign priors to them and let the data sort them out. That is, we sample (e.g., with Markov Chain Monte Carlo \cite{liu2013}) the model parameters $\eta_\mu$ together with $\sigma_\mathrm{out}$ and $P_\mathrm{out}$,
resulting in the joint parameter--hyperparameter posterior probability $p(\boldsymbol{\eta};P_\mathrm{out},\sigma_\mathrm{out}|\mathbf{y})$.
The marginal posterior $p(P_\mathrm{out}|\mathbf{y}) = \int p(\boldsymbol{\eta};P_\mathrm{out},\sigma_\mathrm{out}|\mathbf{y}) \, \mathrm{d}\boldsymbol{\eta} \, \mathrm{d}\sigma_\mathrm{out}$ encodes the fraction of outliers that our scheme identifies in the data, while the probability that each individual datapoint $y_i$ is an outlier is given by
\begin{widetext}
\begin{equation}
\label{eq:piout}
P_{i,\mathrm{out}} = \int 
\frac{
P_\mathrm{out} \times p'(y_i|b_i=1,\boldsymbol{\eta};\sigma_\mathrm{out})
}{
(1 - P_\mathrm{out}) \times p'(y_i|b_i=0,\boldsymbol{\eta}) +
P_\mathrm{out} \times p'(y_i|b_i=1,\boldsymbol{\eta};\sigma_\mathrm{out}) 
} \,
p(\boldsymbol{\eta};P_\mathrm{out},\sigma_\mathrm{out}|y_i) \,
\mathrm{d}\boldsymbol{\eta} \,
\mathrm{d}P_\mathrm{out} \,
\mathrm{d}\sigma_\mathrm{out}.
\end{equation}
%

\paragraph*{Application to pulsar timing.}
Can we apply the mixture scheme to pulsar timing?
The first difficulty that we outlined above for sigma clipping was the necessity of estimating the noise (hyper-)parameters, which may affect the very notion of outlier-ness. But this is no different than what already happens for $\sigma_\mathrm{out}$ and $P_\mathrm{out}$ in the mixture scheme, so we can just sample the noise hyperparameters alongside the other two.

The second difficulty was the requirement of a likelihood that can be factorized into sublikelihoods for each individual point, whereas the most general timing-model likelihood involves a dense vector--matrix products of residuals and noise covariance. There is in fact a form of timing-model likelihood, known as \emph{hierarchical}, which is manifestly factorizable. To explain how it comes about, we begin with the more usual time-correlation form of the likelihood:
%
\begin{equation}
\label{eq:marginalized}
p_\mathrm{GP}(\mathbf{y}|\boldsymbol{\eta}) = \frac{\mathrm{e}^{-\frac{1}{2} \sum_{ij} (y_i - \sum_\mu M_{i\mu} \eta_{\mu}) \bigl(N_{ij} + K_{ij}\bigr)^{-1} (y_j - \sum_\nu M_{j\nu} \eta_{\nu})}}{\sqrt{(2 \pi)^n \det (N + K)}},
\quad
\end{equation}
%
where ``GP'' stands for ``Gaussian-process.''
In this equation, the $y_i$ are the $n$ pulsar-timing residuals; the $\eta_\mu$ are the timing parameters; $M_{i\mu}$ is the design matrix that encodes the effect of changing the timing parameters around their best-fit values; the diagonal matrix $N_{ij} = \delta_{ij} \sigma_i^2$ collects the individual variance of the measurement errors [which are analog to the $\epsilon_i$ of Eq.\ \eqref{eq:leastsquares}]; and the \emph{dense} matrix $K_{ij}$ represents the covariance of correlated noise, a function of a set of noise hyperparameters not shown here to simplify notation.
(See \cite{2014PhRvD..90j4012V} for a review of this formalism.)
Because of $K_{ij}$, the likelihood cannot be factorized.

Recent work (also reviewed in \cite{2014PhRvD..90j4012V}) showed that Eq.\ \eqref{eq:marginalized} for $p_\mathrm{GP}(\mathbf{y}|\boldsymbol{\eta})$ is completely equivalent to the integral of a \emph{hierarchical} likelihood $p_h(\mathbf{y}|\boldsymbol{\eta},\mathbf{c})$:
%
\begin{equation}
\label{eq:hierarchical}
\begin{aligned}
p_\mathrm{GP}(\mathbf{y}|\boldsymbol{\eta}) = \int p_h(\mathbf{y}|\boldsymbol{\eta},\mathbf{c}) \,\mathrm{d}\mathbf{c} &= 
\int
\frac{\mathrm{e}^{-\frac{1}{2} \sum_{i} \bigl(y_i - \sum_a \phi_a(x_i) c_a - \sum_\mu M_{i\mu} \eta_{\mu} \bigr)^2 / \sigma_i^2}}{\sqrt{(2 \pi)^n \prod_i \sigma_i^2}} \times
\frac{e^{-\frac{1}{2} \sum_{ab} c_a (\Phi_{ab})^{-1} c_b}}{\sqrt{(2 \pi)^m \det \Phi}} \, \mathrm{d}\mathbf{c} \\
& = \int \Bigl[
\prod_i p_h(y_i|\boldsymbol{\eta},\mathbf{c}) \Bigr] \times p(\mathbf{c}) \, \mathrm{d}\mathbf{c};
\end{aligned}
\end{equation}
\end{widetext}
here the $m$ basis vectors $\phi_a(x_i)$ reproduce the correlated-noise covariance matrix as $K_{ij} = \sum_{ab} \phi_a(x_i) \Phi_{ab} \phi_b(x_i)$; the $c_a$ are known as the basis \emph{weights}; and the second exponential factor in Eq.\ \eqref{eq:hierarchical} is effectively a Gaussian prior for the weights \cite{rasmussen2006}. (See \cite{2015MNRAS.446.1170V} for a discussion of how the sums over the $\phi_a$ converge to analytical covariance expressions in the case of pulsar timing.)
Thus, we recover Eq.\ \eqref{eq:marginalized} by marginalizing the hierarchical likelihood with respect to the weights, which is also why Eq.\ \eqref{eq:marginalized} is known as the \emph{marginalized} pulsar-timing likelihood.

We see immediately from Eq.\ \eqref{eq:hierarchical} that $p_h(\mathbf{y}|\boldsymbol{\eta},\mathbf{c})$ factorizes with respect to the individual $y_i$, so we can change it into a mixture that accounts for the possibility of outliers. Keeping in mind our picture of timing-model outliers as originating from mistaking noise spikes for pulses, we design a slightly different outlier likelihood than Eq.\ \eqref{eq:pprime}---we posit that TOA outliers are distributed uniformly across a pulsar spin period $P_\mathrm{spin}$. Thus, we make an outlier-tolerant version of Eq.\ \eqref{eq:hierarchical} by way of the simple replacement
\begin{multline}
\label{eq:timingmixture}
p_h(y_j|\boldsymbol{\eta},\mathbf{c}) \, \rightarrow \, p''_h(y_j|\boldsymbol{\eta},\mathbf{c};\sigma_\mathrm{out},P_\mathrm{out}) = \\
(1 - P_\mathrm{out}) \, p_h(y_j|\boldsymbol{\eta},\mathbf{c}) + 
P_\mathrm{out} \frac{1}{P_\mathrm{spin}}.
\end{multline}
\begin{figure*}
\includegraphics[width=\textwidth]{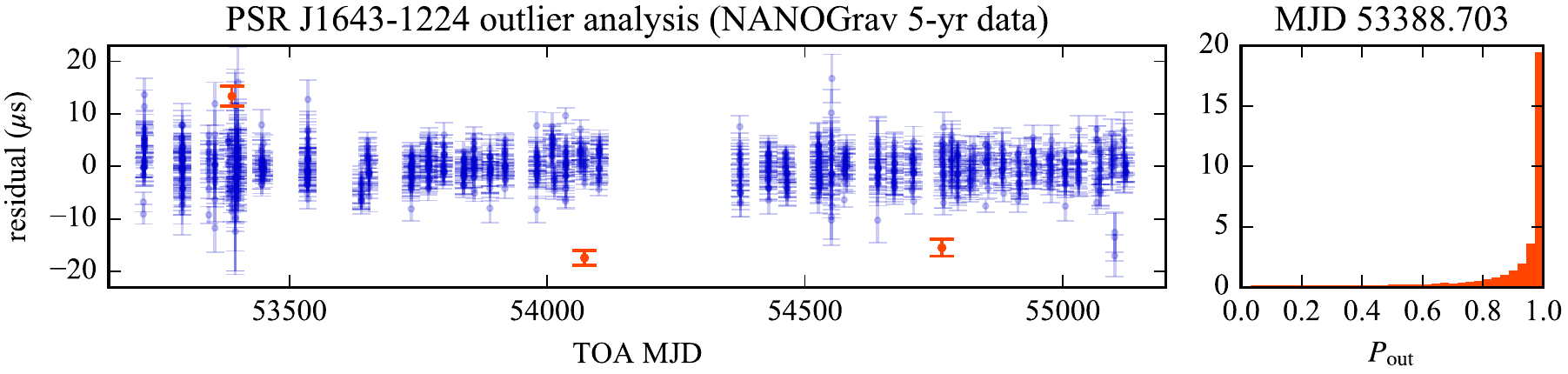}
\vspace{-18pt}
\caption{
Outlier analysis of NANOGrav's 5-year PSR J1643-1224 dataset \cite{2013ApJ...762...94D}. Left: timing residuals, as computed against the best-fit timing-model and noise parameters. The outlier study identifies three strong outliers with $P_{i,\mathrm{out}} \simeq 1$. Right: posterior distribution of $P_{i,\mathrm{out}}$ for the outlier near MJD 53388; in this case $P_{i,\mathrm{out}}$ changes slightly across the posterior distribution of timing-model and noise parameters.
\label{fig:J1643}}
\end{figure*}

Before proceeding with our example, we have three important remarks to make about hierarchical likelihoods in pulsar timing. First, in actual use we never perform the integral over the weights that we wrote in Eq.\ \eqref{eq:hierarchical} [which would get us back to the marginalized likelihood of Eq.\ \eqref{eq:marginalized}], but rather we sample the weights $c_a$ in stochastic fashion, together with the timing-model parameters $\eta_\mu$. So we work with $p_h(\mathbf{y}|\boldsymbol{\eta},\mathbf{c})$ [or, indeed, $p''_h(\mathbf{y}|\boldsymbol{\eta},\mathbf{c};\sigma_\mathrm{out},P_\mathrm{out})$] rather than $p_\mathrm{GP}(\mathbf{y}|\boldsymbol{\eta})$.

Our second remark is a consequence of the first: because the number of sampled parameters increases considerably with the addition of the weights, adopting an efficient stochastic-sampling scheme becomes paramount. Indeed, earlier attempts to use hierarchical likelihoods \cite{2013PhRvD..87j4021L} were stymied by the difficulty of sampling all the weights efficiently, and in particular by \emph{Neal's funnel problem} \cite{neal2003} of sampling each $c_a$ together with the variance-like hyperparameter $\rho_a$ that sets its scale [note that these $\rho_i$ enter Eq.\ \eqref{eq:hierarchical} implicitly through $\Phi$]. 
Recently, in collaboration with J.\ A.\ Ellis, we were able to demonstrate a Hamiltonian sampler \cite{neal2011} with NUTS integration tuning \cite{homan2014}, optimized for pulsar-timing hierarchical likelihoods by a chain of data-aware coordinate transformations \cite{nuts2016}. The transformations come remarkably close to transforming the target distribution into an easily sampled multivariate Gaussian. The sampler is available in the \texttt{Piccard} code at \url{github.com/vhaasteren/piccard}.

Third, unlike Eq.\ \eqref{eq:marginalized}, the hierarchical likelihood involves no inversion of large, dense matrices ($\Phi$ is inverted, but it is usually small and diagonal), so its evaluation is orders of magnitude faster than the evaluation of Eq.\ \eqref{eq:marginalized}, especially for large modern pulsar-timing datasets. In practice, this bonus is partially offset by the larger number of parameters to sample, and by the algebraic manipulations required to tame the target probability distribution. Nevertheless, the \texttt{Piccard} NUTS sampler is remarkably efficient for typical timing-model datasets \cite{nuts2016}.

\paragraph*{Examples.}
The outlier-tolerant hierarchical likelihood, sampled with the \texttt{Piccard} NUTS sampler,\footnote{We obtain 20,000 approximately independent samples, each of which describes the timing-model parameters RAJ, DECJ, PMRA, PMDEC, PX, F0, and F1 (see \cite{2006MNRAS.372.1549E}), as well as the noise hyperparameters EFAC (measurement noise multiplier), EQUAD (quadrature-added noise), the amplitude and exponent of power-law timing noise (represented by 20 sine and cosine Fourier bases), and the outlier hyperparameter $P_\mathrm{out}$.} solves the contamination problem of the dataset in Fig.\ \ref{fig:simulated}: the three outliers are identified as having $P_{i,\mathrm{out}} \simeq 1$ [Eq.\ \eqref{eq:piout}], whereas all other datapoints have $P_{i,\mathrm{out}}$ less than 1\%. Most important, as shown by the dotted histogram in the right panel of Fig.\ \ref{fig:simulated}, the posterior distribution of the EFAC noise parameter becomes unbiased, and tracks closely the posterior obtained by excluding outliers altogether.

Our method can be applied without modification to the real datasets used in pulsar-timing-array searches for gravitational waves. In Fig.\ \ref{fig:J1643} we show the outlier analysis of NANOGrav's 5-year PSR J1643-1224 dataset \cite{2013ApJ...762...94D}, which was completed in $\sim 1$ hour on a recent multicore workstation, again using the \texttt{Piccard} NUTS sampler.\footnote{We obtain 20,000 approximately independent samples, each describing the timing-model parameters RAJ, DECJ, F0, F1, PMRA, PMDEC, PX, PB, A1, XDOT, TASC, EPS1, EPS2, and 40 DMX dispersion-measure parameters (see \cite{2006MNRAS.372.1549E,2013ApJ...762...94D}), as well as the noise hyperparameters EFAC (measurement noise multiplier), EQUAD (quadrature-added noise), ECORR (jitter-like epoch-correlated noise), the amplitude and exponent of power-law timing noise (represented by 20 sine and cosine Fourier bases), and the outlier hyperparameter $P_\mathrm{out}$.}
Three outliers are identified clearly, and shown as the thick red dots and errorbars in the left panel: $P_{i,\mathrm{out}} \simeq 1$ for the TOAs near MJD 54072 and 54765, and slightly less for the TOA near 53388 [viz., $P_{i,\mathrm{out}} = 0.98$, with posterior distribution corresponding to the integrand of Eq.\ \eqref{eq:piout} shown in the right panel]. While these outliers were \emph{not} identified as spurious measurements during the production of the 5-year NANOGrav dataset, our analysis seems rather damning; luckily, the outliers do not significantly affect the estimation of timing-model or noise parameters.

\paragraph*{In conclusion.}

We have described an end-to-end, practical method to identify outliers in pulsar-timing datasets, and to perform outlier-robust statistical inference of timing-model parameters and noise hyperparameters. The treatment of outliers is fully consistent: it accounts for time-correlated timing noise, and for the variation of estimated residuals across the posterior distribution of the noise hyperparameters.

Our method relies crucially on the hierarchical form of the pulsar-timing likelihood [Eq.\ \eqref{eq:hierarchical}], and on the ability to sample it efficiently, which is now possible with a special-purpose Hamiltonian sampler \cite{nuts2016} freely and openly available at \url{github.com/vhaasteren/piccard}. The computational cost of a full inference run scales as $N_\mathrm{pars}^{9/4}$, where $N_\mathrm{pars}$ is the number of sampled parameters. For current NANOGrav datasets \cite{2015ApJ...813...65T,11year}, $N_\mathrm{pars}$ is dominated by the number of jitter-like--noise parameters (one per multi-frequency measurement epoch), which scales linearly with the dataset's timespan. Nevertheless, even datasets with $\sim$ 20,000 TOAs are tractable on workstation-class computers.

Thus, we recommend outlier studies, such as performed above, as a standard step in the production of pulsar-timing-array datasets. Even if a small number of outliers within a large dataset is often tolerated well by non-robust statistical inference, the follow up of strong outlier candidates may reveal undetected problems in radio observations and TOA generation. Indeed, we performed our outlier study in the preparation of the NANOGrav 11-year dataset \cite{11year}. An easily adaptable Python script that performs such a study is available at \url{github.com/vhaasteren/piccard/outliers}.

Our work may be extended in multiple directions. The capability of sampling the hierarchical likelihood efficiently \cite{nuts2016} opens up the possibility of a number of other investigations, such as the characterization of non-Gaussianity (beyond outliers) in timing measurements, similar to what is done in \cite{2014MNRAS.444.3863L}. A mixture probability [Eqs.\ \eqref{eq:pprime} and \eqref{eq:timingmixture}] may also be inserted in other places within the probabilistic model of timing noise. For instance, by modifying the prior for the red-noise weights in Eq.\ \eqref{eq:hierarchical} (which has structure $\propto \exp \{-\mathbf{c}^T \mathsf{\Phi}^{-1} \mathbf{c}/2\}$, with diagonal $\mathsf{\Phi}$) \cite{2014PhRvD..90j4012V} one would provide robustness against quasimonochromatic noise features that may bias the estimation of power-law noise. A similar trick is introduced in \cite{2010PhRvD..82j3007L}.

\vspace{12pt}

\paragraph*{Acknowledgments.}
We thank N.\ Cornish, C.\ Cutler, J.\ Ellis, J.\ Lazio, C.\ Mingarelli, D.\ Nice, and S.\ Taylor for useful comments.
MV was supported by the Jet Propulsion Laboratory RTD program.
RvH was supported by NASA Einstein Fellowship grant PF3-140116. This research was supported in part by National Science Foundation Physics Frontier Center award no. 1430284, and by grant PHYS-1066293 and the hospitality of the Aspen Center for Physics. This work was carried out at the Jet Propulsion Laboratory, California Institute of Technology, under contract to the National Aeronautics and Space Administration. Copyright 2016 California Institute of Technology. Government sponsorship acknowledged.

\newpage

\bibliography{outliers}

\end{document}